\documentstyle[prd,preprint,tighten,aps,eqsecnum,%
amssymb,newlfont]{revtex}
\begin{document}
\preprint{
\begin{tabular}{r}
UWThPh-2002-2\\
August 2002
\end{tabular}
}
\draft
\title{Tests of CPT invariance for \\ 
neutral-flavoured-meson--antimeson mixing}
\author{G.V. Dass}
\address{Physics Department, Indian Institute of Technology \\
\small Powai, Mumbai 400076, India}
\author{W. Grimus}
\address{Institut f\"ur Theoretische Physik, Universit\"at Wien \\
\small Boltzmanngasse 5, A--1090 Wien, Austria}

\maketitle

\begin{abstract}
We focus on two aspects of CPT invariance in
neutral-meson--antimeson ($M^0 \bar M^0$) mixing: 1.\ Tests of CPT
invariance, 
using only the property of ``lack of vacuum regeneration'', 
which occurs as a part of the well-known Lee--Oehme--Yang (LOY) theory;
2.\ Methods for extracting the CPT-violating mixing parameter
$\theta$ through explicit calculations by using the LOY-type theory fully. 
In the latter context, we demonstrate the importance of the C-even
$| M^0 \bar M^0 \rangle$ state. In particular,
by measuring the time dependence of opposite-sign dilepton
events arising from decays of the C-even and C-odd 
$| M^0 \bar M^0 \rangle$ states,
$\theta$ may be disentangled from the parameters 
$\lambda_+$ and $\bar \lambda_-$
characterizing violations of the $\Delta F = \Delta Q$ rule. Furthermore,
these two parameters may also be determined. The same is
true if one uses like-sign dilepton events arising from only the
C-even $| M^0 \bar M^0 \rangle$ state.
\end{abstract}


\newpage

\section{Introduction}

The usual phenomenology of the complex formed by the neutral flavoured
meson $M^0$ ($M^0 = K^0$, $D^0$, $B^0_d$, $B^0_s$) and its antiparticle
$\bar M^0$ is based on the Weisskopf--Wigner approximation (WWA) which
is incorporated into theories of the
Lee--Oehme--Yang (LOY) type \cite{wwa,lee,kabir-book}. 
This complex is investigated extensively for valuable
studies like those of the discrete symmetries CP, T and CPT, 
and of physics beyond the Standard Model 
(for a review see Ref.~\cite{branco}). 
So far, the only
known CP and T non-invariances have arisen in measurements on the
$M^0 \bar M^0$ complex, while CPT conservation is at present consistent 
with all existing data \cite{groom}. 
Therefore, testing CPT invariance at the phenomenological level is 
an important issue (see, e.g., Ref.~\cite{kostelecky}). 
The purpose of this paper is to consider some tests
of CPT invariance in the mixing of $M^0$ and $\bar M^0$, at two
levels \cite{DG} of the WWA.
We will consider the following situations:
\begin{enumerate}
\item
Transitions of single $M^0$ and $\bar M^0$ mesons 
into $M^0$ or $\bar M^0$; this would require flavour-tagging of the initial
\emph{and} final states; 
\item
Transitions of single $M^0$ and $\bar M^0$ mesons into decay channels (e.g.,
$\pi \pi$, $\pi \ell \nu$, $\ldots$); here, only the initial states have to be
tagged for flavour;
\item
Transitions of the C-even and C-odd correlated 
$| M^0 \bar M^0 \rangle$ states into two flavoured mesons $(M^0, \bar 
M^0)$; this would require flavour-tagging of the final states;
\item
Transitions of these correlated states into decay channels, without need for
flavour-tagging. 
\end{enumerate}
We will demonstrate the 
importance of the C-even state \cite{DGL}---particularly for
disentangling CPT violation from violation of the $\Delta F = \Delta Q$ rule 
($F$ means flavour and may be $S$ or $C$ or $B$) in semileptonic decays;
both these violations could arise from physics beyond the Standard Model.
One may note that it is important to allow new physics
through violations of the $\Delta F = \Delta Q$ rule if one is looking
for new physics through CPT violation, especially because of the similarity
\cite{lavoura,xing99} of
the effects of these two types of violations. 

Let us briefly mention some tagging methods.
For neutral kaons, the CPLEAR \cite{CPLEAR} reactions
$\bar p p \to K^+ \pi^- \bar K^0\, /\, K^- \pi^+ K^0$ allow flavour-tagging of
the initial kaon by utilizing
the identity of the accompanying charged kaons and pions. This method is based
only on strangeness conservation in strong interactions. 
Similarly, the reactions \cite{tanner}
$K^0 p \to K^+ n$, $\bar K^0 p \to \pi^+ \Lambda$ (see also
Ref.~\cite{CPLEAR98}) may be used for final-state tagging, also for
decays of the correlated $| K^0 \bar K^0 \rangle$ states. For heavier
flavours, one may also do final-state tagging by using the
flavour-conserving strong interactions---e.g., the ``jet charge
method'' (see, e.g., Ref.~\cite{OPAL}) corresponding to the relevant
flavoured quark. This procedure, used for B mesons,
is a purely empirical procedure of the ``calibrated'' type, wherein, briefly
speaking, one estimates the sign of the charge of the parent flavoured quark
by performing suitable weighted averages over charges of the particle tracks
in the jet produced by the flavoured quark; 
to make the analysis more reliable, the jets
from the parent quark and the parent antiquark are simultaneously
considered. Apart from its empirical nature, the procedure is general.

\section{Outline and formalism}

The WWA is characterized by the introduction of two independently
propagating states $|M_{H,L} \rangle$ which are linear combinations of
the flavour states:
\begin{equation}\label{MHML}
\begin{array}{ll}
| M_H \rangle = p_H | M^0 \rangle + q_H | \bar M^0 \rangle \,,
& |p_H|^2 + |q_H|^2 = 1 \,, \\
| M_L \rangle = p_L | M^0 \rangle - q_L | \bar M^0 \rangle \,,
& |p_L|^2 + |q_L|^2 = 1 \,, 
\end{array}
\end{equation}
where $p_{H,L}$ and $q_{H,L}$ are complex constants. Thus the time
evolution is described by
\begin{equation}\label{theta}
| M_{H,L} \rangle \stackrel{t}{\to} \Theta_{H,L}(t) | M_{H,L} \rangle
\quad \mathrm{with} \quad \Theta_{H,L}(0) = 1 \,,
\end{equation}
where $t$ is the proper time and the $\Theta_{H,L}$ are the
propagation functions. By the same token, a crucial property of the
WWA is the \emph{lack of vacuum regeneration} (called LVR below),
i.e, the absence of 
transitions $| M_{H,L} \rangle \to | M_{L,H} \rangle$ in the time
evolution. Let us define the general probability amplitudes for the
transitions
$| M^0 \rangle \to | M^0 \rangle$,
$| M^0 \rangle \to |\bar M^0 \rangle$,
$| \bar M^0 \rangle \to | M^0 \rangle$ and 
$| \bar M^0 \rangle \to |\bar M^0 \rangle$,
respectively, as $a(t)$, $b(t)$, $\bar b(t)$ and $\bar a(t)$. Then, 
LVR gives \cite{DG}
\begin{eqnarray}
&& \bar b(t) = \alpha b(t) \,,       \label{wwa1b} \\
&& \bar a(t) - a(t) = \beta b(t) \,, \label{wwa1a}
\end{eqnarray}
where $\alpha$ and $\beta$ are complex constants determined in terms of
$p_{H,L}$ and $q_{H,L}$:
\begin{equation}\label{alphabeta}
\alpha = \frac{p_H p_L}{q_H q_L} 
\quad \mbox{and} \quad
\beta = \frac{p_L}{q_L} - \frac{p_H}{q_H} \,.
\end{equation}
Eqs.~(\ref{wwa1b}) and (\ref{wwa1a}) may 
qualitatively be visualized as follows. Using (i) Eq.~(\ref{theta}),
(ii) linearity of the transformation of Eq.~(\ref{MHML}) and its
inverse, and (iii)
the general constraints $\bar a(0) = a(0) = 1$, $b(0) = \bar b(0) = 0$, one
must have $b(t)$, $\bar b(t)$ and $\bar a(t) - a(t)$ all proportional to
$\Theta_H(t) - \Theta_L(t)$.
While $|\alpha| \neq 1$ signifies T non-invariance, $\beta \neq 0$
signifies CPT non-invariance. Using 
Eqs.~(\ref{wwa1b}) and (\ref{wwa1a}), transition rates can be expressed 
as functions of only two amplitudes, say $a$ and $b$; therefore, in
any theory using the LVR (e.g., the LOY theory) 
these equations are often useful in algebraic manipulations.
The remaining part of the WWA can be
expressed as \cite{DG}
\begin{eqnarray}
&&
a + \bar a = \Theta_H + \Theta_L \,,     \label{wwa2a} \\ 
&&
b = q_H q_L (\Theta_H - \Theta_L)/D \,,  \label{wwa2b}
\end{eqnarray}
with $D = p_H q_L + p_L q_H$.

It is useful to define \cite{branco}
\begin{equation}\label{tq}
\theta = \frac{q_H/p_H - q_L/p_L}{q_H/p_H + q_L/p_L}
\quad \mbox{and} \quad
\frac{q}{p}= \sqrt{\frac{q_H q_L}
{p_H p_L}}\, ,
\end{equation}
where both the real and imaginary parts of the 
phase-convention-independent parameter $\theta$
are in principle measurable and violate both CP and CPT;
the quantity $|q/p|-1$ is a measure of CP and T violation in mixing.
Then we can write
\begin{equation}\label{alphabeta1}
\alpha = \left( \frac{p}{q} \right)^2
\quad \mbox{and} \quad
\beta = 2\, \frac{p}{q} \frac{\theta}{\sqrt{1-\theta^2}} \,.
\end{equation}
With the parameters of Eq.~(\ref{tq}), 
the description of the amplitudes $a$, $b$, $\bar a$, $\bar b$ 
in the full WWA is obtained as \cite{branco,kabir-SP}
\begin{equation}
\label{ab}
\begin{array}{lll}
a(t) & = & g_+(t) - \theta g_-(t) \, , \\*[2mm]
b(t) & = & {\displaystyle \frac{q}{p}}\, \sqrt{1-\theta^2} g_-(t) \, , \\*[2mm]
\bar a(t) & = & g_+(t) + \theta g_-(t) \, , \\*[2mm]
\bar b(t) & = & {\displaystyle \frac{p}{q}}\, \sqrt{1-\theta^2} g_-(t) \,,
\end{array}
\end{equation}
where the functions $g_\pm$ are given by
\begin{equation}\label{g+-}
g_\pm(t) = \frac{1}{2} \left( \Theta_H(t) \pm \Theta_L(t) \right) \,.
\end{equation}

So far, the functions $\Theta_{H,L}$ have not been specified. The
exponential decay law of the WWA gives $\Theta_{H,L}$ as
\begin{equation}\label{wwa3}
\Theta_{H,L}(t) = \exp (-it \lambda_{H,L}) 
\quad \mbox{with} \quad
\lambda_{H,L} = m_{H,L} - \frac{i}{2} \Gamma_{H,L} \,,
\end{equation}
where, as usual, $m_{H,L}$ are the real masses and
$\Gamma_{H,L}$ the real decay widths of $| M_{H,L} \rangle$.
In the following we will also need the definitions
$\Delta m = m_H - m_L$ and $\Delta \Gamma = \Gamma_H - \Gamma_L$.
We shall use the expression ``full WWA''
for Eqs.~(\ref{ab}), (\ref{g+-}) and (\ref{wwa3}). 

It is worth remarking that for unknown $\alpha$ and $\beta$,
Eqs.~(\ref{wwa1b}) and (\ref{wwa1a}) provide tests of the LVR property. T (and
CP) invariance gives, in general,
\begin{equation}\label{n}
|b| = |\bar b| \,,
\end{equation}
and CPT (and CP) invariance requires, in general,
\begin{equation}\label{nn}
a = \bar a \,.
\end{equation}
Then, within the LVR, Eq.~(\ref{n}) means $|\alpha| = 1$ and Eq.~(\ref{nn})
means $\beta = 0$. If CPT invariance, viz.\ Eq.~(\ref{nn}) holds, it is not
possible to test the proportionality of $(\bar a - a)$ to $b$, 
which is a characteristic of LVR. 
Thus, CPT invariance within the LVR means Eqs.~(\ref{nn}) and
(\ref{wwa1b}); the characteristic LVR form of Eq.~(\ref{wwa1a}) is then not
relevant. 

In the following sections, we focus on two subjects: 1.\ tests of CPT
invariance within the LVR, 2.\ explicit 
calculations using the full WWA,
with the aim of determining $\theta$. In Section \ref{single},
we consider the one-time transitions described by the four amplitudes 
$a$, $b$, $\bar a$, $\bar b$. 
In Section \ref{single-decay} we summarize, for reference and 
comparison, decays of single
$M^0$ and $\bar M^0$ mesons, which have been extensively investigated;
see, e.g., 
Ref.~\cite{lavoura}. In Section \ref{correlated},
the two-time transitions of the C-even and C-odd correlated $M^0 \bar M^0$
states $| \psi_\pm \rangle$ to $M^0 M^0$, $\bar M^0 \bar M^0$, $M^0 \bar M^0$
and $\bar M^0 M^0$ final states are considered. 
Section \ref{decays} is devoted to the two-time
decays of the correlated states $| \psi_\pm \rangle$ into physical channels
$f$ and $g$. Section \ref{dilepton} deals with explicit calculations by
choosing $f$ and $g$ as semileptonic channels---both like-sign and
opposite-sign dilepton events. Finally, Section \ref{conclusions} gives a
summary. 

\section{Transitions of single mesons 
$M^0$ or $\bar M^0$ to $M^0$, $\bar M^0$}
\label{single}

Let us first consider transitions of single $M^0$ or $\bar M^0$ mesons to
$M^0$, $\bar M^0$, in
analogy to the corresponding T-invariance 
considerations of Refs.~\cite{CPLEAR,kabir70,aharony}.
It has been argued \cite{dass01} that in order to avoid further
assumptions (arising from the use of weak-interaction decays as
substitutes for flavour tags) in the interpretation
\cite{dass01,kabir99,rouge} of data \cite{CPLEAR}, one should directly
measure $|a|$, $|b|$, $|\bar a|$ and $|\bar b|$, and construct
asymmetries out of these. In particular, the experimentally
interesting asymmetries
\begin{equation}\label{KA}
K \equiv \frac{|\bar b|^2 - |b|^2}{|\bar b|^2 + |b|^2}
\quad \mathrm{and} \quad 
A \equiv \frac{|\bar a|^2 - |a|^2}{|\bar a|^2 + |a|^2} 
\end{equation}
test T invariance and CPT invariance, respectively: $K=0$ and
$A=0$. While the LVR relation (\ref{wwa1b}) involving the
time-reversal parameter $\alpha$ gives a clear prediction for $K$,
namely \cite{kabir70,aharony}
\begin{equation}\label{Ka}
K = \frac{|\alpha|^2 - 1}{|\alpha|^2 + 1} = \mathrm{constant}\,,
\end{equation}
the corresponding LVR relation (\ref{wwa1a}) involving the
CPT parameter $\beta$ does not give a simple and testable prediction
for $A$, because $\beta$ and $b$ are not rephasing-invariant
\cite{branco} and the $t$-dependent rephasing-invariant product
$\beta b$ is not easily accessible. However, the LVR relation
(\ref{wwa1a}) may be used to get the bound
\begin{equation}\label{bound}
-|\beta| \leq \frac{|\bar a| - |a|}{|b|} \leq |\beta| \,.
\end{equation}
Unfortunately, this bound is not a clean equality test, in contrast to
Eq.~(\ref{Ka}). 

If we use the full WWA, the appropriate CPT observable $A$ is
obtained as 
\begin{equation}\label{Aa}
A = \frac{|\bar a|^2 - |a|^2}{|\bar a|^2 + |a|^2} =
2\, \mathrm{Re}\, \left[ \theta \frac{g_-(t)}{g_+(t)} \right]
\end{equation}
to first order in the CPT-violating parameter $\theta$.

\section{Decays of single $M^0$, $\bar M^0$ mesons}
\label{single-decay}

We investigate now the decays 
$| M^0(t) \rangle \to | f \rangle$ and 
$| \bar M^0(t) \rangle \to | f \rangle$, where
$| M^0(0) \rangle = | M^0 \rangle$ and 
$| \bar M^0(0) \rangle = | \bar M^0 \rangle$. These have 
the decay rates
\begin{eqnarray}
&& \label{Rt}
R(f,t) =      | \langle f | T | M^0(t) \rangle |^2 = 
\left| a(t) A_f + b(t) \bar A_f \right|^2 \,, \\
&& \label{Rtbar}
\bar R(f,t) = | \langle f | T | \bar M^0(t) \rangle |^2 = 
\left| \bar a(t) \bar A_f + \bar b(t) A_f \right|^2 \,,
\end{eqnarray}
where we have used the definitions
\begin{equation}\label{amp}
\langle f | T |      M^0 \rangle =      A_f \,, \quad
\langle f | T | \bar M^0 \rangle = \bar A_f \,.
\end{equation}
These decays have been discussed previously in the light of CPT
violation in mixing; see, e.g., 
Ref.~\cite{lavoura}. We review them here for comparison with our
alternative method in Section \ref{dilepton}.

In order to exploit the rates (\ref{Rt}) and (\ref{Rtbar}) for the
determination of $\theta$, it is necessary to have information on the
decay amplitudes $A_f$ and $\bar A_f$. Let us focus on 
semileptonic decays with final states 
$X \ell^+ \nu_\ell$ and $\bar X \ell^- \bar\nu_\ell$, where 
$X$ ($\bar X$) is a specific hadronic state.
Allowing for transitions which violate the $\Delta F = \Delta Q$ rule,
we introduce the rephasing-invariant quantities \cite{branco}
\begin{equation}
\label{lambdas}
\lambda_+ = \frac{q}{p}\, \frac{\bar A_+}{A_+}
\quad \mbox{and} \quad
\bar\lambda_- = \frac{p}{q}\, \frac{A_-}{\bar A_-} \, ,
\end{equation}
where $A_+ \equiv A_{\ell^+}$,
$\bar A_- \equiv \bar A_{\ell^-}$,
and so on.
For instance, the CPLEAR Collaboration in Ref.~\cite{CPLEAR01}
considers semileptonic decays of tagged $K^0$ and $\bar
K^0$ with $X = \pi^-$. Using the convenient notation
$R_+(t)$ for having $M^0$ at $t=0$ decaying semileptonically into 
$\ell^+$, etc.,
one obtains \cite{lavoura}
\begin{eqnarray}
R_+(t) & = & |A_+|^2 \left| g_+(t) + g_-(t)
\left( \lambda_+ \sqrt{1-\theta^2} - \theta \right) \right|^2 \,,
\label{Rt+} \\
\bar R_-(t) & = & |\bar A_-|^2 \left| g_+(t) + g_-(t)
\left( \bar\lambda_- \sqrt{1-\theta^2} + \theta \right) \right|^2 \,, 
\label{Rtbar-} \\
R_-(t) & = & |\bar A_-|^2 \left| \frac{q}{p} \right|^2
\left| g_+(t) \bar \lambda_- + g_-(t)
\left( \sqrt{1-\theta^2} - \theta \bar \lambda_- \right) \right|^2 \,,
\label{Rt-} \\
\bar R_+(t) & = & |A_+|^2 \left| \frac{p}{q} \right|^2
\left| g_+(t) \lambda_+ + g_-(t)
\left( \sqrt{1-\theta^2} + \theta \lambda_+ \right) \right|^2 \,.
\label{Rtbar+}
\end{eqnarray}
These four rates allow one to disentangle CPT violation in
mixing from violations of the $\Delta F = \Delta Q$ rule
\cite{lavoura}. 

In order to get a feeling for the experimental results
of Ref.~\cite{CPLEAR01}, it is useful to compare the rates
(\ref{Rt+}), (\ref{Rtbar-}), (\ref{Rt-}), (\ref{Rtbar+}) with the
corresponding ones in Eqs.~(9a)--(9d) of Ref.~\cite{CPLEAR01},
which were expressed there by using a particular rephasing
non-invariant parameterization. One finds the correspondences
$\lambda_+ \leftrightarrow -x$,
$\bar \lambda_- \leftrightarrow -{\bar x}^*$,
$\theta \leftrightarrow 2\delta$,
$(1-|q/p|)/2 \leftrightarrow \mathrm{Re}\, \varepsilon$,
$(1-|\bar A_-/A_+|^2)/4 \leftrightarrow \mathrm{Re}\, y$;
all these supposedly small parameters were retained up to only the
first order.
Note that the short-lived and long-lived kaons
correspond, respectively, to our states 
$| M_L \rangle$ and $| M_H \rangle$. 
Among the three asymmetries constructed out of the four $K^0_{e3}$
decay rates in Ref.~\cite{CPLEAR01}, 
the one relevant for the determination of
$\theta$ is their $A_\delta(t)$, which involves also the complex
parameter $\bar \lambda_- - \lambda_+$. The result of
Ref.~\cite{CPLEAR01} is
\begin{equation}\label{numeric}
\begin{array}{rcl}
\mathrm{Re}\, \theta & = &
(\hphantom{-}6.0 \pm 6.6 \pm 1.2) \times 10^{-4} \,, \\
\mathrm{Im}\, \theta & = &
(-3.0 \pm 4.6 \pm 0.6) \times 10^{-2} \,, \\
\mathrm{Re}\, (\bar \lambda_- - \lambda_+) & = &
(\hphantom{-}0.4 \pm 2.6 \pm 0.6) \times 10^{-2} \,, \\
\mathrm{Im}\, (\bar \lambda_- - \lambda_+) & = &
(\hphantom{-}2.4 \pm 4.4 \pm 0.6) \times 10^{-2} \,,
\end{array}
\end{equation}
where the first error is statistical and the second is systematic.
Though the experimental results (\ref{numeric}) are consistent with
$\theta = 0$ and $\bar \lambda_- - \lambda_+ = 0$, 
the large errors in these results could be concealing sizeable
violations of CPT invariance and of the $\Delta S = \Delta Q$ rule.
For experimental reasons, the full information contained in the four rates
$\stackrel{\scriptscriptstyle (-)}{R}_\pm(t)$ was not accessible, 
and it turns out
that the two complex parameters $\bar \lambda_-$ and $\lambda_+$ were not
fully separated \cite{CPLEAR01}. Though the best ($K^0, \bar K^0$) data 
presently available allow one to determine $\theta$ (with sizeable
errors), they are unable 
to determine separately the $\Delta S = \Delta Q$ rule-violating parameters; 
therefore, heavy ($M^0, \bar M^0$) systems, 
wherein the clean CPLEAR method of flavour-tagging is not
applicable, are likely to pose more severe problems. Consequently, it is
interesting to have---for the purpose of obtaining the above three
complex parameters separately---an alternative procedure which does not
require flavour-tagging. We shall see in Section \ref{dilepton} that
semileptonic decays of C-even correlated states, in addition to those of
the C-odd ones, may provide such an alternative. 

\section{States of two mesons ($M^0$, $\bar M^0$) arising from correlated
states $| M^0 \bar M^0 \rangle$}
\label{correlated}

Let us now consider the entangled states
\begin{equation}
\label{psi}
| \psi_\epsilon \rangle = {\textstyle \frac{1}{\sqrt{2}}}
\left[
| M^0 ( \vec k ) \rangle \otimes | \bar M^0 ( - \vec k ) \rangle 
+ \epsilon
| \bar M^0 ( \vec k ) \rangle \otimes | M^0 ( - \vec k ) \rangle
\right] \,,
\end{equation}
where $\epsilon = \pm 1$ refers to the C-even and C-odd state,
respectively. 
First we discuss probabilities for finding 
$| M^0 (\vec k) \rangle$ (the momentum $\vec k$ pointing to the
left-hand side) at time $t_\ell$ and 
$| M^0 (-\vec k) \rangle$ at time $t_r$ (on the right-hand side), 
etc.\ (see, e.g., Ref.~\cite{dass99}): 
\begin{eqnarray}
P_\epsilon(M^0,t_\ell;M^0,t_r) & = & \frac{1}{2}
\left| a_\ell \bar b_r + \epsilon \bar b_\ell a_r \right|^2 \,, \\
P_\epsilon(\bar M^0,t_\ell;\bar M^0,t_r) & = & \frac{1}{2}
\left| \vphantom{\bar b_r}
b_\ell \bar a_r + \epsilon \bar a_\ell b_r \right|^2 \,, \\
P_\epsilon(M^0,t_\ell;\bar M^0,t_r) & = & \frac{1}{2}
\left| a_\ell \bar a_r + \epsilon \bar b_\ell b_r \right|^2 \,, 
\label{mmb} \\
P_\epsilon(\bar M^0,t_\ell;M^0,t_r) & = & \frac{1}{2}
\left| b_\ell \bar b_r + \epsilon \bar a_\ell a_r \right|^2 \,,
\label{mbm}
\end{eqnarray}
where 
$\stackrel{\scriptscriptstyle (-)}{a}_\ell \equiv 
 \stackrel{\scriptscriptstyle (-)}{a}\!(t_\ell)$ 
and 
$\stackrel{\scriptscriptstyle (-)}{a}_r \equiv 
 \stackrel{\scriptscriptstyle (-)}{a}\!(t_r)$, etc.
No assumption of any discrete symmetry or about the LOY theory and 
WWA is made. 

One may define the asymmetries \cite{dass99}
\begin{eqnarray}
Q_{1\epsilon}(t_\ell,t_r) & = &
\frac{P_\epsilon(M^0,t_\ell;M^0,t_r) - 
      P_\epsilon(\bar M^0,t_\ell;\bar M^0,t_r)}%
     {P_\epsilon(M^0,t_\ell;M^0,t_r) + 
      P_\epsilon(\bar M^0,t_\ell;\bar M^0,t_r)} \,, 
\label{Q1} \\
Q_{2\epsilon}(t_\ell,t_r) & = &
\frac{P_\epsilon(M^0,t_\ell;\bar M^0,t_r) - 
      P_\epsilon(\bar M^0,t_\ell;M^0,t_r)}%
     {P_\epsilon(M^0,t_\ell;\bar M^0,t_r) + 
      P_\epsilon(\bar M^0,t_\ell;M^0,t_r)} \,.
\label{Q2}
\end{eqnarray}
For $\epsilon = -1$, using LVR fully, i.e.,
Eqs.~(\ref{wwa1b}) and (\ref{wwa1a}), one gets
\begin{equation}\label{Q1-}
Q_{1-}(t_\ell,t_r) = \frac{|\alpha|^2-1}{|\alpha|^2+1} = \mbox{constant}\,,
\end{equation}
which equals the one-time asymmetry $K$ of
Eq.~(\ref{Ka}), as previously noted \cite{dass92,khalfin}; neither CPT
invariance nor T invariance has been assumed. 

Using CPT invariance and the LVR property, the
C-even case ($\epsilon = +1$) also gives the result
(\ref{Q1-}). 
Using LVR and then the WWA relations (\ref{ab}) for
calculating $Q_{1+}$, we obtain
\begin{eqnarray}
a_\ell b_r + b_\ell a_r & = &
\frac{q}{p} \sqrt{1 - \theta^2} \left[ g_-(t_\ell+t_r) - 
2 \theta g_-(t_\ell) g_-(t_r) \right] \,, \\
a_\ell b_r + b_\ell a_r + 2 \beta b_\ell b_r & = &
\frac{q}{p} \sqrt{1 - \theta^2} \left[ g_-(t_\ell+t_r) + 
2 \theta g_-(t_\ell) g_-(t_r) \right] \,;
\end{eqnarray}
this gives, 
to first order in symmetry-violating parameters,
\begin{equation}\label{Q1+}
Q_{1+} \simeq \frac{|\alpha|^2-1}{|\alpha|^2+1} -
4\,\mathrm{Re}\, \left[
\frac{\theta\, g_-(t_\ell) g_-(t_r)}{g_-(t_\ell+t_r)} \right] \,.
\end{equation}
The first term on the right-hand side is, of course, identical with
$Q_{1-}$. Therefore, it should be possible to extract $\theta$ from the
time dependence of Eq.~(\ref{Q1+}). In order to get a feeling for the second
term on the right-hand side of Eq.~(\ref{Q1+}), we consider two limiting
cases. For 
$t_\ell, \, t_r \gg |2/\Delta \Gamma|$, one gets
\begin{equation}
Q_{1+} - Q_{1-} \to 2\, \mathrm{Re}\, \theta\: \mathrm{sign}\, (\Delta \Gamma)
\,.
\end{equation}
On the other hand, for small times
$t_\ell, \, t_r \ll 1/\sqrt{(\Delta m)^2 + (\Delta \Gamma/2)^2}$,
one can show that
\begin{equation}
Q_{1+} - Q_{1-} \to 2\, \mathrm{Re}\, \left[ \theta
\left( i \Delta m + \frac{1}{2} \Delta \Gamma \right) \right]
\frac{t_\ell t_r}{t_\ell + t_r} \,.
\end{equation}

Now we come to the asymmetry $Q_{2\epsilon}$. Under the exchange 
$t_\ell \leftrightarrow t_r$, the probabilities (\ref{mmb}) and
(\ref{mbm}) get exchanged, due to invariance under a $180^\circ$ rotation. 
Using Eq.~(\ref{wwa1b}),
one can see that $Q_{2\epsilon}$ is non-zero only if CPT invariance
does not hold, viz.\ $a \neq \bar a$.
This provides a test of CPT invariance within LVR, for both
$\epsilon = \pm 1$: in the probabilities (\ref{mmb}) and
(\ref{mbm}), the part which is odd under $t_\ell \leftrightarrow t_r$
vanishes. 

The LVR relations of Eqs.~(\ref{wwa1b}) and (\ref{wwa1a}) give
\begin{eqnarray}
\lefteqn{P_\epsilon(M^0,t_\ell;\bar M^0,t_r) - 
         P_\epsilon(\bar M^0,t_\ell;M^0,t_r) =} 
\nonumber \\ &&
\mbox{Re}\, \left\{ \left(
a_\ell a_r + \epsilon \alpha b_\ell b_r + 
\beta \frac{1}{2} ( a_\ell b_r + b_\ell a_r ) \right)^* 
( a_\ell b_r - b_\ell a_r ) \, \beta 
\right\} \,. \label{PP}
\end{eqnarray}
This difference (and also $Q_{2\epsilon}$) is non-zero only for $\beta
\neq 0$. Invoking the full WWA gives the asymmetries
$Q_{2\epsilon}$, to first order in $\theta$, as
\begin{eqnarray}
&& Q_{2-} \simeq 2\, \mathrm{Re}\, \left[
\frac{\theta \, G_2(t_\ell,t_r)}{G_1(t_\ell,t_r)} \right] =
-2\, \mathrm{Im} \left[ 
\frac{\theta \sin (t_- \Delta \lambda/2)}{\cos (t_- \Delta \lambda/2)}
\right] \,, \label{Q2-} \\
&& Q_{2+} \simeq 2\, \mathrm{Re}\, \left[
\frac{\theta \, G_2(t_\ell,t_r)}{g_+(t_\ell+t_r)} \right] =
-2\, \mathrm{Im} \left[ 
\frac{\theta \sin (t_- \Delta \lambda/2)}{\cos (t_+ \Delta \lambda/2)}
\right] \,. \label{Q2+}
\end{eqnarray}
Here, we have defined the complex parameter 
$\Delta \lambda = \lambda_H - \lambda_L = \Delta m - i\Delta \Gamma/2$ and
the real parameters $t_\pm = t_\ell \pm t_r$.
Furthermore, $G_1(t_\ell,t_r)$ and $G_2(t_\ell,t_r)$ are, respectively, even
and odd under $t_\ell \leftrightarrow t_r$:
\begin{equation}
G_{\stackrel{\scriptstyle 1}{2}}(t_\ell,t_r) = 
g_+(t_\ell)\, g_\pm(t_r) - g_-(t_\ell)\, g_\mp(t_r) 
= \frac{1}{2} \left[
e^{-i(\lambda_L t_\ell + \lambda_H t_r)} \pm
e^{-i(\lambda_H t_\ell + \lambda_L t_r)} \right] \,.
\end{equation}
Again, one can see that the real and imaginary parts of $\theta$ can be
extracted from measurements of $Q_{2-}$ or $Q_{2+}$.

\section{Decays of the correlated states into physical channels}
\label{decays}

We now come to the decays of $| \psi_\epsilon \rangle$ into the
physical channel $f$ detected at $t_\ell$ and the physical channel $g$
detected at $t_r$. Then the rate is (see, e.g., Ref.~\cite{DG}), 
assuming the closed nature of the 
$[ (|M^0 \rangle, |\bar M^0 \rangle) \leftrightarrow
   (|M_H \rangle, |M_L \rangle) ]$ system,
\begin{eqnarray}
\lefteqn{R_\epsilon(f,t_\ell;g,t_r) =} \nonumber \\
&& \frac{1}{2} \bigg| 
(a_\ell \bar b_r + \epsilon \bar b_\ell a_r ) A_f A_g + 
(b_\ell \bar a_r + \epsilon \bar a_\ell b_r ) \bar A_f \bar A_g + 
\nonumber \\
&&
\left( a_\ell \bar a_r + b_\ell \bar b_r + 
\epsilon ( \bar a_\ell a_r + \bar b_\ell b_r ) \right) \,
\frac{1}{2} (A_f \bar A_g + \bar A_f A_g) + 
\nonumber \\
&& 
\left( a_\ell \bar a_r - b_\ell \bar b_r - 
\epsilon ( \bar a_\ell a_r - \bar b_\ell b_r ) \right) \,
\frac{1}{2} (A_f \bar A_g - \bar A_f A_g) \bigg|^2 \,,
\label{wwa0}
\end{eqnarray}
wherein the transition amplitudes of Eq.~(\ref{amp}) are used.

As for $P_\epsilon(M^0,t_\ell;\bar M^0,t_r)$,
we observe that, in $R_+$, the part which is odd under 
$t_\ell \leftrightarrow t_r$ vanishes if CPT invariance holds within
LVR. Within the full WWA, this
result has been noted earlier in an explicit calculation \cite{DGL};
the present result is based on simpler and more general
considerations. Taking into account both Eqs.~(\ref{wwa1b}) and
(\ref{wwa1a}), the rate $R_+$ is given by
\begin{eqnarray}
\lefteqn{R_+(f,t_\ell;g,t_r) = } \nonumber \\ &&
\frac{1}{2}\, \bigg| (a_\ell b_r + b_\ell a_r) 
\left( \alpha A_f A_g + \bar A_f \bar A_g + 
\beta\, \frac{1}{2} (A_f \bar A_g + \bar A_f A_g) \right) +
\nonumber \\ &&
2 \, b_\ell b_r \left( \beta \bar A_f \bar A_g + 
\alpha\, \frac{1}{2} (A_f \bar A_g + \bar A_f A_g) \right) +
\nonumber \\ &&
a_\ell a_r (A_f \bar A_g + \bar A_f A_g) +
(a_\ell b_r - b_\ell a_r) \, \beta \, 
\frac{1}{2} (A_f \bar A_g - \bar A_f A_g) \bigg|^2 \,.
\label{R+}
\end{eqnarray}
This formula shows that, for $\beta \neq 0$, $R_+$
contains a part odd under $t_\ell \leftrightarrow t_r$.

With the full WWA, the rates $R_\mp$ assume
the well known forms \cite{lavoura}
\begin{eqnarray}
R_-(f,t_\ell;g,t_r) & = & \frac{1}{2}
\left| \vphantom{\frac{q}{p}}
\left[ G_1(t_\ell,t_r) + \theta G_2(t_\ell,t_r) \right]
A_f \bar A_g - \left[ G_1(t_\ell,t_r) - \theta G_2(t_\ell,t_r) \right]
\bar A_f A_g \right. \nonumber \\
& + & \left.
\sqrt{1 - \theta^2} G_2(t_\ell,t_r) \left( \frac{p}{q} A_f A_g
- \frac{q}{p} \bar A_f \bar A_g \right) \right|^2 
\label{R-wwa}
\end{eqnarray}
and \cite{DGL}
\begin{eqnarray}
R_+(f,t_\ell;g,t_r) & = & \frac{1}{2}
\left| \vphantom{\frac{q}{p}}
\left[ g_+(t_+)
- 2 \theta^2 g_- \left( t_\ell \right) g_- \left( t_r \right)
\right] \left( A_f \bar A_g + \bar A_f A_g \right) 
\right. \nonumber \\
& + & \frac{p}{q}\, \sqrt{1 - \theta^2}
\left[ g_- \left( t_+ \right)
- 2 \theta g_- \left( t_\ell \right) g_- \left( t_r \right)
\right] A_f A_g \nonumber \\
& + & \frac{q}{p}\, \sqrt{1 - \theta^2}
\left[ g_- \left( t_+ \right)
+ 2 \theta g_- \left( t_\ell \right) g_- \left( t_r \right)
\right] \bar A_f \bar A_g \nonumber \\
& + & \left.
\theta G_2(t_\ell,t_r) \left( A_f \bar A_g - \bar A_f A_g \right) 
\vphantom{\frac{q}{p}} \right|^2 \,.
\label{R+wwa}
\end{eqnarray}

\section{Dilepton events from correlated decays}
\label{dilepton}

For explicit calculations, we first consider 
opposite-sign dilepton events \cite{xing99,sanda},
i.e., semileptonic decays with $f = X \ell^+ \nu_\ell$
and $g = \bar X \ell^- \bar\nu_\ell$. For illustrating our point, we consider
the same type of lepton on the two sides.
The amplitudes $A_+$, etc.\ and the $\Delta F = \Delta Q$ rule-violating 
parameters $\lambda_+$ and $\bar \lambda_-$ are defined in 
Section \ref{single-decay}. 
We assume that the quantities $\theta$,
$\lambda_+$ and $\bar \lambda_-$,
which describe ``unexpected'' physics,
are small; we retain contributions up to only the first order in these
quantities. 

First, we want to show that, by observing the time dependence of decays into
opposite-sign dilepton events of both 
$| \psi_+ \rangle$ and $| \psi_- \rangle$,
it is possible to disentangle $\theta$, $\lambda_+$,
and $\bar \lambda_-$. This is easily seen by comparing \cite{lavoura}
\begin{equation}\label{R-,+-}
R_-(\ell^+,t_\ell;\ell^-,t_r) = \frac{1}{2} |A_+|^2 |\bar A_-|^2 
\left| G_1(t_\ell,t_r) + 
( \theta - \lambda_+ + \bar \lambda_- ) G_2(t_\ell,t_r) \right|^2
\end{equation}
with \cite{DGL}
\begin{equation}\label{R+,+-}
R_+(\ell^+,t_\ell;\ell^-,t_r) = \frac{1}{2} |A_+|^2 |\bar A_-|^2 
\left| g_+(t_+) + ( \lambda_+ + \bar \lambda_- ) g_-(t_+) +
\theta G_2(t_\ell,t_r) \right|^2 \,.
\end{equation}
The part of the rate $R_-(\ell^+,t_\ell;\ell^-,t_r)$ which is odd
under $t_\ell \leftrightarrow t_r$ determines the combination
$\theta - \lambda_+ + \bar \lambda_-$, whereas in the case of
$R_+(\ell^+,t_\ell;\ell^-,t_r)$ the odd and even parts depend on
$\theta$ and $\lambda_+ + \bar \lambda_-$, respectively.

Considering like-sign dilepton events \cite{lavoura,sanda}, ``new physics''
does not enter 
at first order for the C-odd state \cite{lavoura}:
\begin{equation}\label{R-,++}
R_-(\ell^+,t_\ell;\ell^+,t_r) = 
\frac{1}{2} |A_+|^4 \left| \frac{p}{q} \right|^2 | G_2(t_\ell,t_r) |^2
\end{equation}
and
\begin{equation}\label{R-,--}
R_-(\ell^-,t_\ell;\ell^-,t_r) = 
\frac{1}{2} |\bar A_-|^4 \left| \frac{q}{p} \right|^2 
| G_2(t_\ell,t_r) |^2 \,.
\end{equation}
However, correlated decays of the C-even state into like-sign dilepton
events do contain ``new physics'' at first order:
\begin{equation}\label{R+,++}
R_+(\ell^+,t_\ell;\ell^+,t_r) = \frac{1}{2} |A_+|^4 
\left| \frac{p}{q} \right|^2 \left| g_-(t_+) + 2 \lambda_+ g_+(t_+) -
2\, \theta g_-(t_\ell) g_-(t_r) \right|^2 
\end{equation}
and
\begin{equation}\label{R+,--}
R_+(\ell^-,t_\ell;\ell^-,t_r) = \frac{1}{2} |\bar A_-|^4 
\left| \frac{q}{p} \right|^2 \left| g_-(t_+) + 2 \bar\lambda_- g_+(t_+) +
2\, \theta g_-(t_\ell) g_-(t_r) \right|^2 \,.
\end{equation}
From these two rates, which are obviously symmetric under 
$t_\ell \leftrightarrow t_r$, the quantities $\theta$, $\lambda_+$ and
$\bar \lambda_-$ could be disentangled because the
functions of $t_\ell$ and $t_r$ with which they are associated are different.

A remark is now in order concerning the comparison of the formulas of this
section with experiment.
In general, the amplitudes $A_\pm$, $\bar A_\pm$ will depend on
the detailed configuration of the final state $X \ell^+ \nu_\ell$ or 
$\bar X \ell^- \bar\nu_\ell$,
i.e., on the particle content of $X$ ($\bar X$)
and the momenta and polarizations
of all particles in the final states. Let us denote the sum over various
choices of $X$ ($\bar X$) and various configurations of spins and momenta
detected on the left-hand side by $\langle \ldots \rangle_\ell$
and the corresponding sum detected on the right-hand side by 
$\langle \ldots \rangle_r$.
Consider, as an example, the rate
$R_-(\ell^+,t_\ell;\ell^+,t_r)$. Taking into consideration the
summation over the final configurations, we obtain (see also
Refs.~\cite{CPLEAR01,wu})
\begin{eqnarray}
\lefteqn{ \langle R_-(\ell^+,t_\ell;\ell^+,t_r) \rangle_{\ell,r} =} 
\nonumber \\
&&
\frac{1}{2} \Bigg\{ | G_2(t_\ell,t_r) |^2 \left| \frac{p}{q} \right|^2
\langle |A_+|^2 \rangle_\ell\, \langle |A_+|^2 \rangle_r + 
\nonumber \\
&& \left.
2\, \mathrm{Re}\, \left[
G_2(t_\ell,t_r)^* G_1(t_\ell,t_r) \left( \frac{p}{q} \right)^* \left( 
\langle |A_+|^2 \rangle_\ell\, \langle A_+^* \bar A_+ \rangle_r - 
\langle A_+^* \bar A_+ \rangle_\ell\, \langle |A_+|^2 \rangle_r
\right) \right] \right\} =
\nonumber \\
&& 
\frac{1}{2} \left| \frac{p}{q} \right|^2
\langle |A_+|^2 \rangle_\ell\, \langle |A_+|^2 \rangle_r
\left| G_2(t_\ell,t_r) + G_1(t_\ell,t_r) ( \lambda_+^r -
\lambda_+^\ell ) \right|^2 \,, \label{R+av}
\end{eqnarray}
where at most the first order in the small parameters $\theta$ and
\begin{equation}\label{la+}
\lambda_+^r = \frac{q}{p}\,
\frac{\langle A_+^* \bar A_+ \rangle_r}{\langle |A_+|^2 \rangle_r} \,,
\quad
\lambda_+^\ell = \frac{q}{p}\,
\frac{\langle A_+^* \bar A_+ \rangle_\ell}{\langle |A_+|^2 \rangle_\ell} 
\end{equation}
has been retained. We similarly obtain
\begin{equation}
\langle R_-(\ell^-,t_\ell;\ell^-,t_r) \rangle_{\ell,r} = 
\frac{1}{2} \left| \frac{q}{p} \right|^2
\langle |\bar A_-|^2 \rangle_\ell\, \langle |\bar A_-|^2 \rangle_r
\left| G_2(t_\ell,t_r) + G_1(t_\ell,t_r) ( \bar\lambda_-^r -
\bar\lambda_-^\ell ) \right|^2 \,, \label{R-av}
\end{equation}
where 
\begin{equation}\label{la-}
\bar\lambda_-^r = \frac{p}{q}\,
\frac{\langle \bar A_-^* A_- \rangle_r}{\langle |\bar A_-|^2 \rangle_r} \,,
\quad
\bar\lambda_-^\ell = \frac{p}{q}\,
\frac{\langle \bar A_-^* A_- \rangle_\ell}{\langle |\bar A_-|^2 \rangle_\ell} 
\,.
\end{equation}
The ratio of the rates in Eqs.~(\ref{R+av}) and (\ref{R-av}) has a
constant value if the $\Delta F = \Delta Q$ rule holds, in which
case the lepton charge \emph{is} the flavour tag; if, in addition,
CPT invariance in the amplitudes holds and if for a given side 
(viz.\ left or right), the states and configurations
summed over in Eqs.~(\ref{R+av}) and (\ref{R-av}) are CPT-conjugates
of each other, the constant value 
is just the time-reversal parameter $|p/q|^4$; see
Ref.~\cite{dass92} for corresponding remarks if tagging of the final
flavoured mesons is not replaced by their semileptonic decays.
On the other hand, one now sees that,
in $\langle R_-(\ell^+,t_\ell;\ell^+,t_r) \rangle_{\ell,r}$ and 
$\langle R_-(\ell^-,t_\ell;\ell^-,t_r) \rangle_{\ell,r}$, 
violations of the $\Delta F = \Delta Q$ rule cancel if left- and
right-hand sides are summed over identical states and configurations.
Implicitly, we have made this assumption of identical left and right
summations in all our results in Eqs.~(\ref{R-,+-})--(\ref{R+,--}), where 
the $\Delta F = \Delta Q$ rule-violating parameters
$\lambda_+$ and $\bar\lambda_-$ should be perceived as the effective
parameters of Eqs.~(\ref{la+}) and (\ref{la-}), respectively (of course,
now we have $\lambda_+^r = \lambda_+^\ell$ and 
$\bar\lambda_-^r = \bar\lambda_-^\ell$).
Eqs.~(\ref{R+av}) and (\ref{R-av}) illustrate this point for 
Eqs.~(\ref{R-,++}) and (\ref{R-,--}), respectively, and show the importance of
identical left and right summations.

\section{Conclusions}
\label{conclusions}

In this paper we have discussed two items. Firstly, we have proposed
tests of CPT invariance within only the
\emph{lack-of-vacuum-regeneration} (LVR) property. This means testing
$a = \bar a$ and $\bar b \propto b$ together (see Eqs.~(\ref{nn}) and
(\ref{wwa1b})). The second item is the determination, by assuming the full WWA,
of the parameter $\theta$ of Eq.~(\ref{tq}), which is a measure of CPT
violation in $M^0 \bar M^0$ mixing.
In the following, subscripts $\mp$ refer to the C-odd and C-even 
$| M^0 \bar M^0 \rangle$ states $| \psi_\mp \rangle$ of
Eq.~(\ref{psi}), respectively. 

As for the first point, we have noted the following qualitative tests:
\begin{enumerate}
\renewcommand{\theenumi}{\roman{enumi}}
\item
The asymmetry $Q_{1+}(t_\ell,t_r)$ (see Eq.~(\ref{Q1})) equals
$Q_{1-}(t_\ell,t_r)$ of 
Eq.~(\ref{Q1-}); these are asymmetries for transitions
into $M^0 M^0$ and $\bar M^0 \bar M^0$ final states.
\item
The asymmetries $Q_{2\mp}(t_\ell,t_r)$ for $M^0 \bar M^0$ and 
$\bar M^0 M^0$ final states vanish. Correspondingly, the 
($t_\ell \leftrightarrow t_r$)-odd parts of the probabilities
(\ref{mmb}) and (\ref{mbm}) vanish.
\item
The ($t_\ell \leftrightarrow t_r$)-odd part of the decay rate
$R_+(f,t_\ell;g,t_r)$ of Eq.~(\ref{R+}) vanishes.
\end{enumerate}

The second item of our paper, viz.\ methods for the determination of
$\theta$, involves explicit computations of observables within the
full WWA. 
These observables include the cases where $\theta = 0$
would reproduce one of the above-mentioned tests, i.e., 
$Q_{1+}$ in Eq.~(\ref{Q1+}), $Q_{2-}$ in Eq.~(\ref{Q2-}), 
$Q_{2+}$ in Eq.~(\ref{Q2+}) and $R_+$ in Eq.~(\ref{R+wwa}). 
In addition, we have the rate $R_-$ in Eq.~(\ref{R-wwa}) and,
for one-time single meson transitions, the asymmetry $A$ in Eq.~(\ref{Aa}). Of
these six observables which involve 
$\theta$, the decay rates $R_\pm$ involve also unknown decay amplitudes
and, therefore, cannot be directly used for the determination of CPT
violation in mixing.

In view of the previous difficulties \cite{DGL,lavoura,xing99} 
in achieving this last goal by using the decay rates $R_\pm$ for correlated
decays 
of the C-even and C-odd $| M^0 \bar M^0 \rangle$ states, 
we have further investigated semileptonic decays in this context.
We have shown in Section \ref{dilepton} (see also Ref.~\cite{DGL}) 
that the three complex parameters $\theta$, $\lambda_+$ and
$\bar\lambda_-$, where the latter two quantities parameterize
violations of the $\Delta F = \Delta Q$ rule, may be separately
determined either by comparing the time dependence of opposite-sign
dilepton events from the state $| \psi_- \rangle$ with that from
$| \psi_+ \rangle$, or by considering both possible charges in the
like-sign dilepton events from $| \psi_+ \rangle$ alone.
Note that, if one wishes to determine $\theta$ alone, it is sufficient
to consider the time dependence of only
$R_+(\ell^+,t_\ell;\ell^-,t_r)$ \cite{DGL}.
The disentanglement of the above-mentioned three parameters is---in
principle---possible also by using semileptonic decays of single 
mesons $M^0$ and $\bar M^0$ (see Ref.~\cite{lavoura} and Section
\ref{single-decay}); however, that requires initial-state
tagging. As shown in Section \ref{single-decay}, 
by considering the best presently available data \cite{CPLEAR01}, it is
useful to have an alternative procedure which does not require
flavour-tagging. Our proposal for considering dilepton events from the decays
of $| \psi_- \rangle$ \emph{and} $| \psi_+ \rangle$ may provide such an
alternative. 

Though some of the experiments proposed in this paper 
require difficult steps like flavour-tagging and 
a study of the decay of the C-even $M^0 \bar M^0$
state $| \psi_+ \rangle$ \cite{ways}, the importance of testing the 
fundamental property of CPT invariance may make the effort 
worthwhile.

\end{document}